\documentclass[pra,twocolumn,showpacs,superscriptaddress,floatfix, reprint]{revtex4}

\usepackage[OT1]{fontenc}
\usepackage[usenames,dvipsnames]{color}
\usepackage[latin1]{inputenc}
\usepackage[english]{babel}
\usepackage{graphicx}
\usepackage{color}
\usepackage[Gray,squaren]{SIunits}
\usepackage{xspace}
\usepackage{upgreek}
\usepackage{ulem}
\usepackage{epstopdf}
\usepackage{amssymb,amsmath,verbatim,ulem}
\newcommand{\ket}[1]{|{#1}\rangle}
\newcommand{\bra}[1]{\langle{#1}|}

\begin{document}

\title{Single reference atomic based MW interferometry using EIT}
\author{Dangka Shylla and Kanhaiya Pandey}
\affiliation{Department of Physics, Indian Institute of Technology Guwahati, Guwahati, Assam 781039, India }
 \email{kanhaiyapandey@iitg.ernet.in}


%\affiliation{
%\affiliation{
%\affiliation{%


\date{\today{}}

%\begin{abstract} 
%In this work, we theoretically study a scheme to develop an atomic based MW interferometry, which in the contrast to previously studied system requires only single reference instead of  three. The atomic based MW interferometry bypasses the conventional electrical circuit based MW interferometry in superior fashion i.e. the performance is not limited by the Nyquist thermal noise and having much higher bandwidth. 
%This scheme involves magnetic sublevels in the Rydberg states and hence will be suitable in the even isotope of  Yb or alkaline earth elements such as Mg, Ca, Sr, where there is no complicacy due to absence of the hyperfine levels and it is easy to address the magnetic sublevels. Further the wavelength of the two lasers to excite the Rydberg state, are very close, which cancels the Doppler shift more effectively in the counter-propagating configuration and amplitude/phase sensitivity increases significantly. We characterize this system for the amplitude/phase sensitivity of the unknown MW field in terms of the reference MW field and compare it to the previously studied systems.
%\end{abstract}

\begin{abstract} 

Recently atomic based MW electrometry is experimentally demonstrated and interferometry has been proposed. The proposed interferometry bypasses the conventional, electrical circuit based MW interferometry in much superior fashion. However, this scheme requires three different references for characterizing the unknown MW field. In this work we theoretically study a scheme to develop an atomic based MW interferometry having only one referenced MW field. This scheme involves magnetic sublevels in the Rydberg states and hence will be suitable in even isotope of Yb or alkaline earth element where there is no complicacy due to absence of the hyperfine levels. Further, the wavelengths to excite the Rydberg states, are very close and hence cancels the Doppler shift more effectively which increases the amplitude sensitivity. We characterize this system for the phase and the amplitude of the unknown MW field w.r.t to the known field and compare it to the previously studied systems.

%We characterize this system for the amplitude/phase sensitivity of the unknown MW field in terms of the reference MW field and compare it to the previously studied systems.

%This scheme involves magnetic sublevels of the Rydberg states and hence will be suitable in the even isotope of  Yb or alkaline earth elements such as Mg, Ca, Sr, where there is no complicacy due to absence of the hyperfine levels and it is easy to address the magnetic sublevels.

\end{abstract}

%\pacs{42.50.Md, 42.25.Dd}

\maketitle

%\tableofcontents

%Below is the brief description about the work, which is going to open a new field i.e. atomic based MW interferometry replacing the conventional electrical circuit in much superior fashion.
%\section{Introduction}

\section{Introduction}
Atomic based standards and measurements have gained lot of reliability and is already stablished for time and length due to its accuracy, precision, and reproducibility \cite{HAL06}. Atomic based DC and AC (MW and RF) magnetometry is in use at the device level, due to it's impressive sensitivity and spatial resolutions \cite{BUR07,LRD13,HDM13,HGP15,CLD15}. However, there are many physical quantities which are yet to be standardized based on the atom. The Micro-wave (MW) field is one of them. The characterization of the MW field is very important and has immediate applications in the communication and radar technologies specially in active sensing and synthetic aperture \cite{SAR98}. The MW field is generally characterized by the electrical circuit based MW interferometry whose performance is greatly limited by Nyquist thermal noise and the bandwidth of the circuit \cite{KIY81,IVT09,ITW98}. Recently there was a great boost towards characterization of the MW based upon the atom \cite{KSB10,SSK12} utilizing the very high electric polarizability of available closely space Rydberg states. However there was no progress towards atomic based MW interferometry i.e. complete characterization. This is because previously studied system was insensitive to the phase of the MW fields. In the effort of atomic based MW interferometry recently a loopy ladder system has been proposed \cite{SOP18} in Rb and is expected to be \textbf{two orders} of magnitude more sensitive as compared to the experimentally demonstrated MW electrometry \cite{KSB10,SSK12}. However, the proposed MW interferometry \cite{SOP18} requires three reference MW fields.
In this work, we theoretically study a double loopy ladder-system realized in Yb using the magnetic sublevels to propose single reference MW interferometry. The scheme is based upon the interference between the two sub-system causing transparency of probe which has phase dependency of unknown MW field w.r.t the reference field. This scheme is more suitable in the even isotope of  Yb or alkaline earth elements such as Mg, Ca, Sr, where there is no complicacy due to absence of the hyperfine levels and it is easy to address the magnetic sublevels.
Further, the wavelength of the two lasers to excite the Rydberg state, are very close, which cancels the Doppler shift more effectively in the counter-propagating configuration and amplitude sensitivity increases significantly. We characterize this system for the phase and the amplitude of the unknown MW field w.r.t to the known field and compare it to the previously studied systems.

\section{Model System}
\begin{figure}
   \begin{center}
      \includegraphics[width =8cm]{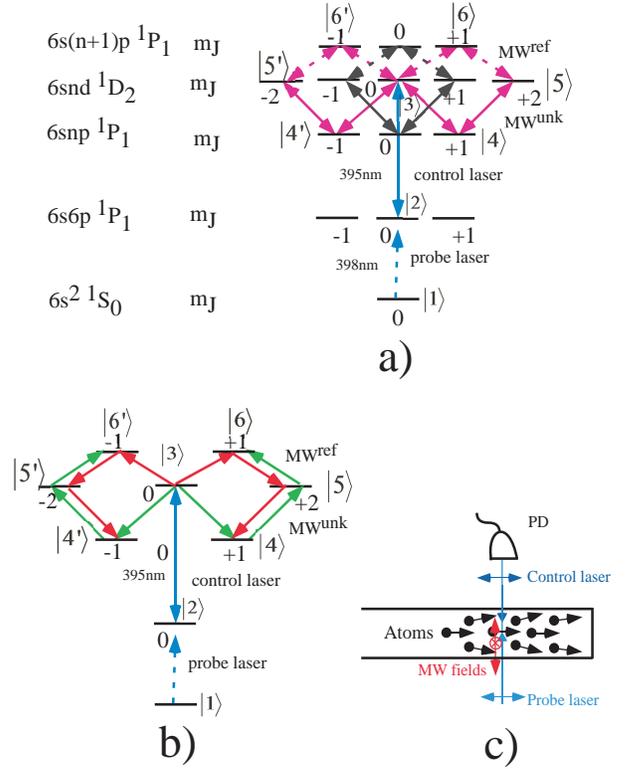}
      \caption{(Color online). (a) The energy level diagram for double loopy ladder system for single reference MW interferometry in Yb. 
 (b) Transitions shown by green and red arrow lines are the two sets of sub-systems closing the loop. (c) Schematic representation of the experimental set up to realize the double loopy ladder system. It consists of the atomic source, laser beams and MW fields.}
      \label{CLladdersys}
   \end{center}
\end{figure}

For our study we choose a double loopy ladder system in even isotope of Yb as shown in  Fig. \ref{CLladdersys}a. This scheme is also valid for even isotope of earth alkaline element such Sr, Ca and Mg which has similar level structure. The transitions from the ground state, 6s$^2$ $^1$S$_0$ to the first excited singlet state, 6s6p $^1$P$_1$ and from the first excited singlet state, 6s6p $^1$P$_1$ to the Rydberg state, 6snd $^1$D$_2$ are driven by the probe laser at wavelength 398~nm and control laser at wavelength 395~nm respectively. The other transitions from the Rydberg singlet D state 6snd $^1$D$_2$ to another Rydberg singlet P state, 6s (n-1)p $^1$P$_1$  and from the Rydberg singlet Rydberg D state 6snd $^1$D$_2$ to Rydberg singlet P state, 6s np $^1$P$_1$ are driven by the unknown MW field and the reference MW field respectively. The higher Rydberg states of Yb has been theoretically calculated and measured experimentally \cite{VIL81, XXH94}. The bandwidth of this interferometry can range from MHz ( n $\sim$150), GHz ( n $\sim$ 100), few tens of GHz(n $\sim$  60) to THz(n $\sim$ 10) \cite{VIL81, XXH94}. 

We choose the quantization axis along the control and probe laser polarization direction and hence these two lasers drive the $\pi$ transition. The polarization of the unknown and the reference MW field is perpendicular to the quantization axis and are decomposed into $\sigma^+$ and $\sigma^-$ polarization. The relevant transition driven by the optical and MW fields are shown in Fig. \ref{CLladdersys}a.   
 
The AC electric field interacting with the atomic system corresponding to the transition  $\ket{i}$ $\rightarrow$ $\ket{j}$ is $E_{ij}e^{i(\omega_{ij}t+\phi_{ij})}$, where $E_{ij}$ is the amplitude at frequency, $\omega_{ij}$ and $\phi_{ij}$ is the phase. $\Omega_{ij} =-d_{ij}E_{ij}e^{i\phi_{ij}}/\hbar$ is the Rabi frequency associated with the electric field that couples the  $\ket{i}$ $\rightarrow$ $\ket{j}$ transition having dipole moment matrix element  $d_{ij}$. Therefore, we define $\Omega_{12}$ and $\Omega_{23}$ to be the Rabi frequencies of the probe and the control field respectively and $\Omega^{\textrm{unk}}_{34}$, $\Omega^{\textrm{unk}}_{34'}$, $\Omega^{\textrm{unk}}_{45}$, $\Omega^{\textrm{unk}}_{4'5'}$, $\Omega^{\textrm{ref}}_{56}$, $\Omega^{\textrm{ref}}_{5'6'}$, $\Omega^{\textrm{ref}}_{36}$ and $\Omega^{\textrm{ref}}_{3'6'}$  to be the Rabi frequencies of the control MW fields. Note that the subscript here denotes the transition driven by them whereas, the superscript \lq{unk\rq} and \lq{ref\rq} refers to the unknown and the reference MW fields  respectively. After incorporating the Clebsch Gorden coefficients for the transitions driven by the MW field and the decomposition of the linearly polarized MW electric fields into $\sigma^+$ and $\sigma^-$ we have the following relations \small{} $|\Omega^{\textrm{unk}}_{45}|=\sqrt{6}|\Omega^{\textrm{unk}}_{34}|$=$|\Omega^{\textrm{unk}}_{4'5'}|=\sqrt{6}|\Omega^{\textrm{unk}}_{34'}|=\frac{1}{\sqrt{2}}|\Omega^{\textrm{unk}}|$,  \normalsize{} where \small{}$|\Omega^{\textrm{unk}}|$  \normalsize{} is the magnitude of the maximum Rabi frequency associated with the transition 6snd $^1$D$_2$ $\rightarrow$ 6snp $^1$P$_1$ transition. Similarly, for the reference MW field we have \small{}$|\Omega^{\textrm{ref}}_{56}|=\sqrt{6}|\Omega^{\textrm{ref}}_{36}|$ =$|\Omega^{\textrm{ref}}_{5'6'}|=\sqrt{6}|\Omega^{\textrm{ref}}_{3'6'}|$= $\frac{1}{\sqrt{2}}|\Omega^{\textrm{ref}}|$.

 \normalsize{}The schematic representation for the experimental setup of phase dependent MW interferometry is as shown in Fig.\ref{CLladdersys}(c) in which a probe laser at 398~nm and a control laser at 395~nm are counter-propagating inside the Yb atomic beam. It is hard to make a glass cell for the alkaline earth element or Yb in this case, as the sublimation temperature of these elements is around 700~K. Hence, the atomic beam is a good option and has already been used previously\cite{DBB05,PSK09,YPP15}. The typical divergence of a roughly collimated atomic beam corresponds to a transverse temperature of around 1~K. We will be dealing with calculations for 1~K and also in the extreme case of temperature at 700~K where there is no collimation of the atomic beam.

The total Hamiltonian for this system in the dipole moment approximation can be written as
%\begin{align}
%&H=\bigg[\sum^5_{i=1,4',5'}\frac{\hbar\Omega_{i,i+1}}{2}\left(e^{i\omega_{i,i+1}t}+e^{-i\omega_{i,i+1}t}\right)\ket{i}\bra{i+1}+\frac{\hbar\Omega_{36}}{2}\left(e^{i\omega_{36}t}+e^{-i\omega_{36}t}\right)\ket{3}\bra{6}\\\nonumber
%&+\frac{\hbar\Omega_{36'}}{2}\left(e^{i\omega_{36'}t}+e^{-i\omega_{36'}t}\right)\ket{3}\bra{6'}+h.c.\Big]+\sum_{i=1,4',5',6'}^{6}\hbar\omega_i\ket{i}\bra{i}\\\nonumber
%\end{align}

\begin{align}
H&=\bigg[\sum^5_{i=1}\frac{\hbar\Omega_{i,i+1}}{2}\left(e^{i\omega_{i,i+1}t}+e^{-i\omega_{i,i+1}t}\right)\ket{i}\bra{i+1}
\nonumber\\&
+\sum^{5'}_{j=4'}\frac{\hbar\Omega_{j,j+1}}{2}\left(e^{i\omega_{j,j+1}t}+e^{-i\omega_{j,j+1}t}\right)\ket{j}\bra{j+1}\nonumber\\&
+\frac{\hbar\Omega_{36}}{2}\left(e^{i\omega_{36}t}+e^{-i\omega_{36}t}\right)\ket{3}\bra{6}
\nonumber\\&
+\frac{\hbar\Omega_{36'}}{2}\left(e^{i\omega_{36'}t}+e^{-i\omega_{36'}t}\right)\ket{3}\bra{6'}+h.c.\Big]
\nonumber\\&
+\sum_{i=1}^{6}\hbar\omega_i\ket{i}\bra{i}+\sum_{j=4'}^{{6'}}\hbar\omega_j\ket{j}\bra{j}\\\nonumber
\end{align}
where, $\hbar\omega_i$ and $\hbar\omega_j$ are energies of the states $\ket{i}$ and $\ket{j}$ respectively.

%Hence, the total Hamiltonian will be
%\begin{equation}
%H=H_0+H_I
%\end{equation}
In the rotating frame with rotating wave approximation the above Hamiltonian can be written as
\begin{align}
H&=\hbar[0\ket{1}\bra{1}-\delta_{12}\ket{2}\bra{2}-(\delta_{12}+\delta_{23})\ket{3}\bra{3}
\nonumber\\&
-(\delta_{12}+\delta_{23}-\delta_{34})\ket{4}\bra{4}-(\delta_{12}+\delta_{23}-\delta_{34'})\ket{4'}\bra{4'}
\nonumber \\&
-(\delta_{12}+\delta_{23}-\delta_{34}+\delta_{45})\ket{5}\bra{5}-(\delta_{12}+\delta_{23}-\delta_{34'}+\delta_{4'5'})\ket{5'}\bra{5'}
\nonumber\\&
-(\delta_{12}+\delta_{23}-\delta_{34}+\delta_{45}+\delta_{56})\ket{6}\bra{6}
\nonumber \\&
-(\delta_{12}+\delta_{23}-\delta_{34'}+\delta_{4'5'}+\delta_{5'6'})\ket{6'}\bra{6'}
\nonumber\\&
+\frac{\Omega_{12}}{2}\ket{1}\bra{2}+\frac{\Omega_{23}}{2}\ket{2}\bra{3}
+\frac{\Omega_{34}}{2}\ket{3}\bra{4}+\frac{\Omega_{34'}}{2}\ket{3}\bra{4'} 
\nonumber\\ &
+\frac{\Omega_{45}}{2}\ket{4}\bra{5}+\frac{\Omega_{45}}{2}\ket{4'}\bra{5'}
+\frac{\Omega_{56}}{2}\ket{5}\bra{6}+\frac{\Omega_{5'6'}}{2}\ket{5'}\bra{6'} 
\nonumber\\&
+\frac{\Omega_{36}}{2}e^{i(\delta_{34}-\delta_{45}-\delta_{56}+\delta_{36})t}\ket{3}\bra{6}
\nonumber\\&
+\frac{\Omega_{36'}}{2}e^{i(\delta_{34'}-\delta_{4'5'}-\delta_{5'6'}+\delta_{36'})t}\ket{3}\bra{6'}+h.c.\Big]
\end{align}
where, $\delta_{12}=\omega^L_{12}-(\omega_2-\omega_1)$, $\delta_{23}=\omega^L_{23}-(\omega_3-\omega_2)$  are the detunings of the probe and control lasers, $\delta_{34}=\omega^L_{34}-(\omega_3-\omega_4)$, $\delta_{45}=\omega^L_{45}-(\omega_5-\omega_4)$, $\delta_{56}=\omega_{56}-(\omega_6-\omega_5)$, $\delta_{36}=\omega_{36}-(\omega_6-\omega_3)$  $\delta_{34'}=\omega^L_{34'}-(\omega_3-\omega_{4'})$, $\delta_{4'5'}=\omega^L_{4'5'}-(\omega_{5'}-\omega_{4'})$, $\delta_{5'6'}=\omega_{5'6'}-(\omega_{6'}-\omega_{5'})$ and $\delta_{36'}=\omega_{36}-(\omega_{6'}-\omega_3)$ are the detunings for the MW fields for the respective transitions. Note that the Hamilitonian $H$ is time dependent except for a particular condition when $\delta_{34}-\delta_{45}-\delta_{56}+\delta_{36}=0$ and $\delta_{34'}-\delta_{4'5'}-\delta_{5'6'}+\delta_{36'}=0$.  
To investigate the dynamics of the double loopy ladder atomic system, we employ the density matrix approach using Liouville's equation. This equation gives the time evolution of the density matrix, $\rho$ as $\dot{\rho}=\frac{i}{\hbar}[H, \rho]-\frac{1}{2}\{\Gamma,\rho\}$ where, $\Gamma$ is the relaxation matrix. The advantage of using this equation is the fact that it contains both statistical as well as quantum mechanical information about the system which on solving, yields the following set of differential equations:

\begin{align}
\dot{\rho}_{12}&=i\frac{\Omega_{12}}{2}{(\rho_{11}-\rho_{22})}+i\frac{\Omega_{23}^*}{2}{\rho_{13}}-\gamma_{12}\rho_{12}\nonumber\\
\dot{\rho}_{13}&=-\gamma_{13}\rho_{13}-i\frac{\Omega_{12}}{2}{\rho_{23}}+i\frac{\Omega_{23}}{2}{\rho_{12}}+i\frac{{\Omega_{34}}^*}{2}{\rho_{14}}\nonumber\\&
+i\frac{{\Omega_{34'}}^*}{2}{\rho_{14'}}+i\frac{{\Omega_{36}}^*}{2}e^{-i(\delta_{34}-\delta_{45}-\delta_{56}+\delta_{36})t}{\rho_{16}}\nonumber\\&
+i\frac{{\Omega_{36}}^*}{2}e^{-i(\delta_{34'}-\delta_{4'5'}-\delta_{5'6'}+\delta_{36'})t}{\rho_{16'}}\nonumber\\
\dot{\rho}_{14}&=-\gamma_{14}\rho_{14}-i\frac{\Omega_{12}}{2}{\rho_{24}}+i\frac{\Omega^{\textrm{ref}}_{34}}{2}{\rho_{13}}+i\frac{{\Omega^{\textrm{ref}}_{45}}^*}{2}{\rho_{15}}\nonumber\\
\dot{\rho}_{14'}&=-\gamma_{14'}\rho_{14'}-i\frac{\Omega_{12}}{2}{\rho_{24'}}+i\frac{\Omega_{34'}}{2}{\rho_{13}}+i\frac{{\Omega_{4'5'}}^*}{2}{\rho_{15'}}\nonumber\\
\dot{\rho}_{15}&=-\gamma_{15}\rho_{15}-i\frac{\Omega_{12}}{2}{\rho_{25}}+i\frac{\Omega_{45}}{2}{\rho_{14}}+i\frac{{\Omega_{56}}^*}{2}{\rho_{16}}\nonumber\\
\dot{\rho}_{15'}&=-\gamma_{15'}\rho_{15'}-i\frac{\Omega_{12}}{2}{\rho_{25'}}+i\frac{\Omega_{4'5'}}{2}{\rho_{14'}}+i\frac{{\Omega_{5'6'}}^*}{2}{\rho_{16'}}\nonumber\\
\dot{\rho}_{16}&=-\gamma_{16}\rho_{16}-i\frac{\Omega_{12}}{2}{\rho_{26}}+i\frac{\Omega_{36}}{2}e^{i(\delta_{34}-\delta_{45}-\delta_{56}+\delta_{36})t}{\rho_{13}}
\nonumber\\&
+i\frac{{\Omega_{56}}}{2}{\rho_{15}}\nonumber\\
\dot{\rho}_{16'}&=-\gamma_{16'}\rho_{16'}-i\frac{\Omega_{12}}{2}{\rho_{26'}}
\nonumber\\&
+i\frac{\Omega_{36'}}{2}e^{i(\delta_{34'}-\delta_{4'5'}-\delta_{5'6'}+\delta_{36'})t}{\rho_{13}}+i\frac{{\Omega_{5'6'}}}{2}{\rho_{15'}}
\end{align}
Where,
\newline
$\gamma_{12}=\left[\gamma^{dec}_{12}-i\delta_{12}\right]$, 
\\
$\gamma_{13}=\left[\gamma^{dec}_{13}-i\left(\delta_{12}+\delta_{23}\right)\right]$, 
\\
$\gamma_{14}=\left[\gamma^{dec}_{14}-i\left(\delta_{12}+\delta_{23}-\delta_{34}\right)\right]$, 
\\
$\gamma_{14'}=\left[\gamma^{dec}_{14'}-i\left(\delta_{12}+\delta_{23}-\delta_{34'}\right)\right]$, 
\\
$\gamma_{15}=\left[\gamma^{dec}_{15}-i\left(\delta_{12}+\delta_{23}-\delta_{34}+\delta_{45}\right)\right]$,
\\
$\gamma_{15'}=\left[\gamma^{dec}_{15'}-i\left(\delta_{12}+\delta_{23}-\delta_{34'}+\delta_{4'5'}\right)\right]$, 
\\
$\gamma_{16}=\left[\gamma^{dec}_{16}-i\left(\delta_{12}+\delta_{23}-\delta_{34}+\delta_{45}+\delta_{56}\right)\right]$ and 
\\
$\gamma_{16'}=\left[\gamma^{dec}_{16'}-i\left(\delta_{12}+\delta_{23}-\delta_{34'}+\delta_{4'5'}+\delta_{5'6'}\right)\right]$.
\\
 $\gamma^{dec}_{ij}=\left(\frac{\Gamma_i+\Gamma_j}{2}\right)$ is decoherence rate between level $\ket{i}$ and $\ket{j}$, $\Gamma_i$ and $\Gamma_{j}$ is the total decay rates of states $\ket{i}$ and $\ket{j}$. \\

For our calculations, we take the value of  $\gamma^{dec}_{12}$ which is the decoherence rate between level $\ket{1}$ and $\ket{2}$ to be ~$2\pi\times$14 MHz. $\gamma^{dec}_{12}$ is mainly dominated by the natural radiative decay of excited state 6s6p $^1$P$_1$, $\Gamma_2$ which is found to be $2\pi\times$28~MHz. We also take $\gamma^{dec}_{13}=\gamma^{dec}_{14}=\gamma^{dec}_{14'}=\gamma^{dec}_{15}=\gamma^{dec}_{15'}=\gamma^{dec}_{16}=\gamma^{dec}_{16'}=\gamma^{dec}=2\pi\times$100kHz and these decoherences are mainly dominated by the laser linewidths of the probe and control lasers wavelength as compared to the radiative decay rate which is  ~2$\pi\times$1~kHz of the Rydberg states\cite{SSK13}.

The above equations can be solved by using the weak probe approximation under the steady state condition i.e. $\dot{\rho_{ij}}=0$ for all $i$ and $j$. In the case of weak probe approximation, there will be no population transfer and hence the time evolution of the population  terms i.e. the diagonal terms of the density matrix  can be approximated as $\rho_{11}\approx1$,  $\rho_{22}\approx$ $\rho_{33}\approx$ $\rho_{44}\approx$ $\rho_{55}\approx$ $\rho_{66}\approx$ $\rho_{4'4'}\approx$ $\rho_{5'5'}\approx$ $\rho_{6'6'}\approx0$. The off-diagonal terms as $\rho_{ij}$=$\rho_{ji}\approx0$  for $i=2 $, $ j=3,4,4',5,5',6,6'$; $i=3$, $ j=4,4',5,5',6,6'$; $i=4$, $ j=5,5',6,6'$ and $i=5 $, $ j=6,6'$. After insertion of the approximations in steady state for the above set of differential equations from the density matrix, we obtain the following new set of linear algebraic equations:

\begin{align}
\rho_{12}&=\frac{i}{2}\frac{\Omega_{12}}{\gamma_{12}}+\frac{i}{2}\frac{\Omega^*_{23}}{\gamma_{12}}\rho_{13}\nonumber\\
\rho_{13}&=\frac{i}{2}\frac{\Omega_{23}}{\gamma_{13}}\rho_{12}+\frac{i}{2}\frac{{\Omega_{34}}^*}{\gamma_{13}}\rho_{14}+\frac{i}{2}\frac{{\Omega_{34'}}^*}{\gamma_{13}}\rho_{14'}+\frac{i}{2}\frac{{\Omega_{36}}^*}{\gamma_{13}}\rho_{16}\nonumber\\&+\frac{i}{2}\frac{{\Omega_{36}}^*}{\gamma_{13}}\rho_{16'}\nonumber\\
\rho_{14}&=\frac{i}{2}\frac{\Omega_{34}}{\gamma_{14}}\rho_{13}+\frac{i}{2}\frac{{\Omega_{45}}^*}{\gamma_{14}}\rho_{15}\nonumber\\
\rho_{14'}&=\frac{i}{2}\frac{\Omega_{34'}}{\gamma_{14'}}\rho_{13}+\frac{i}{2}\frac{{\Omega_{4'5'}}^*}{\gamma_{14'}}\rho_{15'}\nonumber\\
\rho_{15}&=\frac{i}{2}\frac{\Omega_{45}}{\gamma_{15}}\rho_{14}+\frac{i}{2}\frac{{\Omega_{56}}^*}{\gamma_{15}}\rho_{16}\nonumber\\
\rho_{15'}&=\frac{i}{2}\frac{\Omega_{4'5'}}{\gamma_{15'}}\rho_{14'}+\frac{i}{2}\frac{{\Omega_{5'6'}}^*}{\gamma_{15'}}\rho_{16'}\nonumber\\
\rho_{16}&=\frac{i}{2}\frac{\Omega_{36}}{\gamma_{16}}\rho_{13}+\frac{i}{2}\frac{\Omega_{56}}{\gamma_{16}}\rho_{15}\nonumber\\
\rho_{16'}&=\frac{i}{2}\frac{\Omega_{36'}}{\gamma_{16'}}\rho_{13}+\frac{i}{2}\frac{\Omega_{5'6'}}{\gamma_{16'}}\rho_{15'}
\label{aprcpl}
\end{align}

The density matrix element, $\rho_{12}$   is potentially related to the refractive index n of the probe laser as $n=1+3\lambda_p^2N/(2\pi)(\Gamma_2/\Omega_{12})\rho_{12}$ in which $\lambda_p=398~nm$ is the wavelength of the probe laser and N is atomic number density\cite{BLJ95,KYP14}. In order to establish an analytical formulation for $\rho_{12}$  which is directly proportional to the absorption experienced by the probe field, we solve the above linear algebraic equations. The above equations gives solution for $\rho_{12}$ as

\begin{align}
&\rho_{12}=\frac{\frac{i}{2}\frac{\Omega_{12}}{\gamma_{12}}}{1+\frac{\frac{1}{4}\frac{|\Omega_{23}|^2}{\gamma_{12}\gamma_{13}}}{1+\textrm{\textcolor{red}{EITATA1}}+\textrm{\textcolor{green}{EITATA2}}+\textrm{Int}+\textrm{\textcolor{red}{EITATA1'}}+\textrm{\textcolor{green}{EITATA2'}}+\textrm{Int'}}}\\
&\textrm{Where,} \nonumber \\
&\textrm{\textcolor{green}{EITATA1}}=\frac{\frac{1}{4}\frac{|\Omega^{\textrm{unk}}_{34}|^2}{\gamma_{13}\gamma_{14}}}{1+\frac{\frac{1}{4}\frac{|\Omega^{\textrm{unk}}_{45}|^2}{\gamma_{14}\gamma_{15}}}{1+\frac{1}{4}\frac{|\Omega^{\textrm{ref}}_{56}|^2}{\gamma_{15}\gamma_{16}}}};\textrm{\textcolor{red}{EITATA2}}=\frac{\frac{1}{4}\frac{|\Omega^{\textrm{ref}}_{36}|^2}{\gamma_{13}\gamma_{16}}}{1+\frac{\frac{1}{4}\frac{|\Omega^{\textrm{ref}}_{56}|^2}{\gamma_{15}\gamma_{16}}}{1+\frac{1}{4}\frac{|\Omega^{\textrm{unk}}_{45}|^2}{\gamma_{14}\gamma_{15}}}};\\
&\textrm{\textcolor{green}{EITATA1'}}=\frac{\frac{1}{4}\frac{|\Omega^{\textrm{unk}}_{34'}|^2}{\gamma_{13}\gamma_{14'}}}{1+\frac{\frac{1}{4}\frac{|\Omega^{\textrm{unk}}_{4'5'}|^2}{\gamma_{14'}\gamma_{15'}}}{1+\frac{1}{4}\frac{|\Omega^{\textrm{ref}}_{5'6'}|^2}{\gamma_{15'}\gamma_{16'}}}};\textrm{\textcolor{red}{EITATA2'}}=\frac{\frac{1}{4}\frac{|\Omega^{\textrm{ref}}_{36'}|^2}{\gamma_{13}\gamma_{16'}}}{1+\frac{\frac{1}{4}\frac{|\Omega^{\textrm{ref}}_{5'6'}|^2}{\gamma_{15'}\gamma_{16'}}}{1+\frac{1}{4}\frac{|\Omega^{\textrm{unk}}_{4'5'}|^2}{\gamma_{14'}\gamma_{15'}}}};\\
&\textrm{Int}=-\frac{\frac{1}{8}\frac{|\Omega^{\textrm{unk}}_{34}||\Omega^{\textrm{unk}}_{45}||\Omega^{\textrm{ref}}_{56}||\Omega^{\textrm{ref}}_{36}|cos(\phi)}{\gamma_{13}\gamma_{14}\gamma_{15}\gamma_{16}}}{1+\frac{1}{4}\frac{|\Omega^{\textrm{unk}}_{45}|^2}{\gamma_{14}\gamma_{15}}+\frac{1}{4}\frac{|\Omega^{\textrm{unk}}_{56}|^2}{\gamma_{15}\gamma_{16}}};\\
%\phi=\phi^{\textrm{ref}}_{36}-\phi^{\textrm{unk}}_{34}-\phi^{\textrm{unk}}_{45}-\phi^{\textrm{ref}}_{56}\\
&\textrm{Int'}=-\frac{\frac{1}{8}\frac{|\Omega^{\textrm{unk}}_{34'}||\Omega^{\textrm{unk}}_{4'5'}||\Omega^{\textrm{ref}}_{5'6'}||\Omega^{\textrm{ref}}_{36'}|cos(\phi)}{\gamma_{13}\gamma_{14'}\gamma_{15'}\gamma_{16'}}}{1+\frac{1}{4}\frac{|\Omega^{\textrm{unk}}_{4'5'}|^2}{\gamma_{14'}\gamma_{15'}}+\frac{1}{4}\frac{|\Omega^{\textrm{unk}}_{5'6'}|^2}{\gamma_{15'}\gamma_{16'}}};
%\phi=\phi^{\textrm{ref}}_{36'}-\phi^{\textrm{unk}}_{34'}-\phi^{\textrm{unk}}_{4'5'}-\phi^{\textrm{ref}}_{5'6'}
\label{anasol}
\end{align}
 where, $\phi=\phi^{\textrm{ref}}_{36}-\phi^{\textrm{unk}}_{34}-\phi^{\textrm{unk}}_{45}-\phi^{\textrm{ref}}_{56}=\phi^{\textrm{ref}}_{36'}-\phi^{\textrm{unk}}_{34'}-\phi^{\textrm{unk}}_{45'}-\phi^{\textrm{ref}}_{56'}$=2($\phi^{\textrm{ref}}-\phi^{\textrm{unk}}$).
In order to verify the approximation made above, we have checked the analytical solution of $\rho_{12}$ given by the Eq. [5] and the complete numerical solution in the steady state for various values of control fields and detunings. It has excellent agreement between complete numerical and approximated analytical solution.

\section{Results and Discussions}
 In this section, we analyze the probe absorption in presence of the control laser and the MW fields.  
As shown in Fig. \ref{CLladdersys}a, the unknown and the reference MW field forms three closed loops, out of which two loops $\ket{3}$ $\leftrightarrow$$\ket{4}$ $\leftrightarrow$ $\ket{5}$ $\leftrightarrow$ $\ket{6}$ $\leftrightarrow$$\ket{3}$ and $\ket{3}$$\leftrightarrow$$\ket{4'}$ $\leftrightarrow$$\ket{5'}$$\leftrightarrow$$\ket{6'}$$\leftrightarrow$ $\ket{3}$, represented with magenta color are connected to the control laser and hence contribute to the phase sensitive modification of the absorption of the probe laser. The loop represented with black color is not connected with control laser and hence it is idle for the probe absorption. The first closed loop, $\ket{3}$$\leftrightarrow$$\ket{4}$$\leftrightarrow$$\ket{5}$$\leftrightarrow$$\ket{6}$$\leftrightarrow$$\ket{3}$  can be realized by two sub-systems $\ket{3}$$\rightarrow$$\ket{4}$$\rightarrow$$\ket{5}$$\rightarrow$$\ket{6}$ and $\ket{3}$$\rightarrow$$\ket{6}$$\rightarrow$$\ket{5}$$\rightarrow$$\ket{4}$ shown with red and green arrows respectively sharing a common $\ket{1}$$\rightarrow$$\ket{2}$$\rightarrow$$\ket{3}$ ladder system. Similarly, the second loop $\ket{3}$$\leftrightarrow$$\ket{4'}$$\leftrightarrow$$\ket{5'}$$\leftrightarrow$$\ket{6'}$$\leftrightarrow$$\ket{3}$ can be also realized by two sub-systems $\ket{3}$$\rightarrow$$\ket{4'}$$\rightarrow$$\ket{5'}$$\rightarrow$$\ket{6'}$ and $\ket{3}$$\rightarrow$$\ket{6'}$$\rightarrow$$\ket{5'}$$\rightarrow$$\ket{4'}$ shown with red and green arrows respectively  and sharing the same $\ket{1}$$\rightarrow$$\ket{2}$$\rightarrow$ $\ket{3}$ ladder system as shown in  Fig. \ref{CLladdersys}b.

 To realize the functions of various control fields, we activate them one by one in the following sequence. But let us consider only the first loop and the second loop later on as it is evident that the same process occurs in the second loop.  The control laser, $\Omega_{23}$ causes reduction in the absorption of the probe laser, $\Omega_{12}$ and is known as EIT. For the path shown with the green arrows, the control field, $\Omega^{\textrm{unk}}_{34}$ recovers the absorption against the EIT and  is known as EITA \cite{PAN13}. In a similar way, the control fields $\Omega^{\textrm{unk}}_{45}$ and $\Omega^{\textrm{ref}}_{56}$ causes EITAT and EITATA \cite{PAN13} expressed by \textcolor{green}{EITATA1}  in Eq. [5]. The other path shown with red arrows will also cause EITATA but by sequence of the control fields $\Omega^{\textrm{ref}}_{36}$, $\Omega^{\textrm{ref}}_{56}$, $\Omega^{\textrm{unk}}_{45}$  and $\Omega^{\textrm{unk}}_{34}$, which is expressed by \textcolor{red}{EITATA2} in Eq. [5]. The term int in the expression of $\rho_{12}$ corresponds to the interference between the two sub-systems causing \textcolor{green}{EITATA1} and \textcolor{red}{EITATA2} and is phase $\phi$ dependent. Similarly, in the second loop the path shown with green arrows by sequence of control fields $\Omega^{\textrm{unk}}_{34'}$, $\Omega^{\textrm{unk}}_{4'5'}$, $\Omega^{\textrm{ref}}_{5'6'}$ and $\Omega^{\textrm{ref}}_{36'}$ causes \textcolor{green}{EITATA1$'$} in Eq. [5]. Also, in other path shown with red arrows will cause \textcolor{red}{EITATA2$'$} by sequence of the control fields $\Omega^{\textrm{ref}}_{36'}$,  $\Omega^{\textrm{ref}}_{5'6'}$,  $\Omega^{\textrm{unk}}_{4'5'}$  and $\Omega^{\textrm{unk}}_{34'}$ in Eq. [5]. Likewise, the term int$'$ in the expression of $\rho_{12}$ corresponds to the interference between the two sub-systems causing \textcolor{green}{EITATA1$'$} and \textcolor{red}{EITATA2$'$} and is phase $\phi$ dependent as well.  

\begin{figure}
   \begin{center}
      \includegraphics[width =\linewidth]{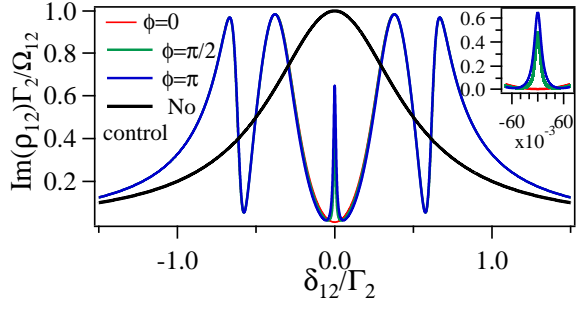}
      \caption{(Color online). Normalized absorption (Im($\rho_{12}$)$\Gamma_2/\Omega_{12}$) vs $\delta_{12}/\Gamma_2$ of the probe laser with $\delta_{23}=\delta_{34}=\delta_{34'}=\delta_{45}=\delta_{4'5'}=\delta_{56}=\delta_{5'6'}=\delta_{36}=\delta_{36'}=0$, $|\Omega_{23}|=\Gamma_2$, $|\Omega^{\textrm{ref}}|=\sqrt{2}\Gamma_2$  and $|\Omega^{\textrm{unk}}|=\sqrt{2}$$\times$$0.1\Gamma_2$. The inset is a zoomed absorption profile around the central peak.}
      \label{Abs_various_phase}
   \end{center}
\end{figure}

 We define the normalized absorption [$(\Gamma_2/\Omega_{12})$Im($\rho_{12})$] i.e. for the stationary atoms the absorption of the probe laser at resonance in the absence of all the control lasers  is 1 as shown by the peak of the black curve in Fig.\ref{Abs_various_phase}. First, we investigate the normalized absorption  (Im($\rho_{12}$)$\Gamma_2/\Omega_{12}$) vs probe detuning ($\delta_{12}$) for three different phases, $\phi=0, \pi/2 $ and $ \pi$ as shown in Fig. \ref{Abs_various_phase}. The double loopy ladder systems $\ket{1}$$\rightarrow$$\ket{2}$$\leftrightarrow$$\ket{3}$$\leftrightarrow$$\ket{4}$$\leftrightarrow$$\ket{5}$$\leftrightarrow$$\ket{6}$ and $\ket{1}$$\rightarrow$$\ket{2}$$\leftrightarrow$$\ket{3}$$\leftrightarrow$$\ket{4'}$$\leftrightarrow$$\ket{5'}$$\leftrightarrow$ $\ket{6'}$ are symmetric with overlapping absorption peaks for $\delta_{14}=\delta_{14'}$,  $\delta_{15}=\delta_{15'}$,  $\delta_{16}=\delta_{16'}$, $\Omega^{\textrm{unk}}_{34}=\Omega^{\textrm{unk}}_{34'}$, $\Omega^{\textrm{unk}}_{45}=\Omega^{\textrm{unk}}_{4'5'}$, $\Omega^{\textrm{ref}}_{36}=\Omega^{\textrm{ref}}_{36'}$ and $\Omega^{\textrm{ref}}_{56}=\Omega^{\textrm{ref}}_{5'6'}$. For $\phi=0$ at line center of the probe absorption, the two sets of sub system causing \textcolor{green}{EITATA1}  and \textcolor{red}{EITATA2} or \textcolor{green}{EITATA1$'$} and \textcolor{red}{EITATA2$'$} interfere destructively with each other and there is transparency. But for $\phi=\pi$, the two sets of sub systems causing \textcolor{green}{EITATA1}  and \textcolor{red}{EITATA2} or \textcolor{green}{EITATA1$'$} and \textcolor{red}{EITATA2$'$}  interfere constructively with each other and there is maximum absorption.  
% For the central absorption peak i.e. at $\delta_{12}=0$, only the linewidth depends upon the phase but not the position, while both the position and the linewidth depends upon the phase ($\phi$) for the other four absorption peaks.

For large Rabi frequencies of the control laser and MW fields, the absorption peaks are well separated. This separation, the linewidth and the amplitude of these can be well understood using the dressed state approach. For general control fields detunings and Rabi frequencies, the position of the absorption peaks (dressed states) will be complicated. However, the expression becomes straighforward at zero detunings of the control field and the MW fields. \normalsize{}The central dressed state is a superposition of the bare atomic states and is expressed as \tiny{} $\frac{1}{\sqrt{A}}$\Bigg[$\frac{-e^{i\phi}|\Omega^{\textrm{ref}}_{36}||\Omega^{\textrm{unk}}_{45}| +|\Omega^{\textrm{unk}}_{34}| |\Omega^{\textrm{ref}}_{56}|}{|\Omega_{23}||\Omega^{\textrm{unk}}_{45}|}$$\ket{2}$-$\frac{|\Omega^{\textrm{ref}}_{56}|}{|\Omega^{\textrm{unk}}_{45}|}\ket{4}$+$\ket{6}$\Bigg], \normalsize{}where \tiny{} $A=$\Bigg[$\Bigg|\frac{-e^{i\phi}|\Omega^{\textrm{ref}}_{36}||\Omega^{\textrm{unk}}_{45}| +|\Omega^{\textrm{unk}}_{34}| |\Omega^{\textrm{ref}}_{56}|}{|\Omega_{23}||\Omega^{\textrm{unk}}_{45}|}$$\Bigg|^2$+$\frac{|\Omega^{\textrm{ref}}_{56}|^2}{|\Omega^{\textrm{unk}}_{45}|^2}$+$1$\Bigg].\normalsize{} Its linewidth is given by \tiny{}$\frac{1}{A}$\Bigg[$\Bigg|\frac{-e^{i\phi}|\Omega^{\textrm{ref}}_{36}||\Omega^{\textrm{unk}}_{45}| +|\Omega^{\textrm{unk}}_{34}| |\Omega^{\textrm{ref}}_{56}|}{|\Omega_{23}||\Omega^{\textrm{unk}}_{45}|}$$\Bigg|^2$$\Gamma_2$+$\frac{|\Omega^{\textrm{ref}}_{56}|^2}{|\Omega^{\textrm{unk}}_{45}|^2}$$\Gamma_4$+$\Gamma_6$\Bigg] \normalsize{}which depends on the phase. The amplitude of the peak is proportional to \tiny{}$\frac{1}{A}$$\Bigg|\frac{-e^{i\phi}|\Omega^{\textrm{ref}}_{36}||\Omega^{\textrm{unk}}_{45}| +|\Omega^{\textrm{unk}}_{34}| |\Omega^{\textrm{ref}}_{56}|}{|\Omega_{23}||\Omega^{\textrm{unk}}_{45}|}$$\Bigg|^2$\normalsize{}=\tiny{}$\frac{(1-cos\phi)\frac{|\Omega^{\textrm{ref}}|^2}{3|\Omega_{23}|^2}}{(1-cos\phi)\frac{|\Omega^{\textrm{ref}}|^2}{3|\Omega_{23}|^2}+\frac{|\Omega^{\textrm{ref}}|^2}{|\Omega_{\textrm{unk}}|^2}+1}$\normalsize{}. From this expression it is clear that the probe absorption is zero at $\phi=0$ and is maximum at $\phi=\pi$. 

\begin{figure}
   \begin{center}
      \includegraphics[width =\linewidth]{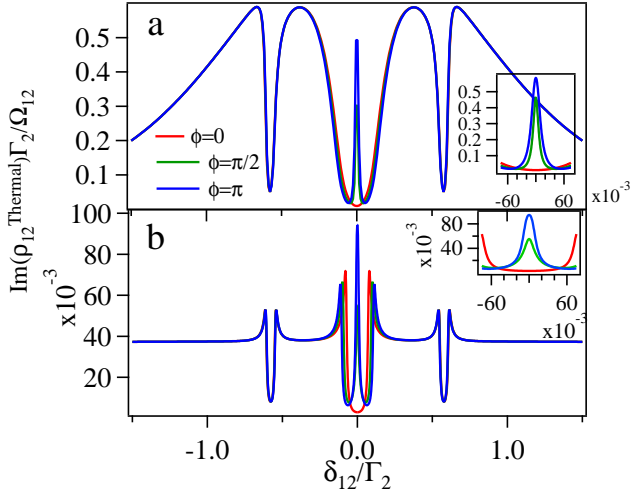}
      \caption{(Color online). Normalized absorption of the probe laser with thermal averaging (Im($\rho^{\textrm{Thermal}}_{12}$)$\Gamma_2/\Omega_{12}$) vs $\delta_{12}/\Gamma_2$ of the probe laser with $\delta_{23}=\delta_{34}=\delta_{34'}=\delta_{45}=\delta_{4'5'}=\delta_{56}=\delta_{5'6'}=\delta_{36}=\delta_{36'}=0$, $|\Omega_{23}|=\Gamma_2$, $|\Omega^{\textrm{ref}}|=\sqrt{2}\Gamma_2$  and $|\Omega^{\textrm{unk}}|= $ $\sqrt{2}\times0.1\Gamma_2$.  (a) T=1~K (b) T=700~K. The insets are zoomed absorption profile around the central peak of (a) and (b). }
      \label{abs_detuning_Dopp}
   \end{center}
\end{figure}
Now, we investigate the effect of temperature on the absorption profile considering the atomic beam to be divergent. The thermal averaging of $\rho_{12}$ is done numerically at two temperatures i.e. at T=1~K and at T=700~K for the counter-propagating configuration of the probe ($\Omega_{12}$) and the control lasers ($\Omega_{23}$) with wave vectors $k_{398}$ and $k_{395}$ respectively,  by replacing $\delta_{12}$ with $\delta_{12}+k_{398}v$ and $\delta_{23}$ with $\delta_{23}-k_{395}v$ for moving atoms with velocity $v$, while the Doppler shift for the MW fields are ignored. Further, the $\rho_{12}$ is weighted by the Maxwell Boltzman velocity distribution function and integrated over the velocity as $\rho^{\textrm{Thermal}}_{12}=\sqrt{\frac{m}{2\pi k_B T}}\int\rho_{12}(v)e^{-\frac{mv^2}{2k_B T}}dv$, where $k_B$ is Boltzman constant and $m$ is the atomic mass of Yb. The integration is done over velocity range which is two times of $\sqrt{\frac{k_BT}{m}}$. Unlike the previously studied system in Rb \cite{SOP18} the two-photon Doppler mismatch for the probe and the control lasers here is very small in comparison to $\Gamma_2$ as the wavelength of two optical transition are very close due to which there is no broadening of the central absorption peak as shown in Fig. \ref{abs_detuning_Dopp} by thermal averaging. The narrowing of the EIT window and the enhanced absorption at the wing is still observed which has been extensively studied previously \cite{KPW05,PWN08,IKN08,PKP16}.   

\begin{figure}
   \begin{center}
      \includegraphics[width =\linewidth]{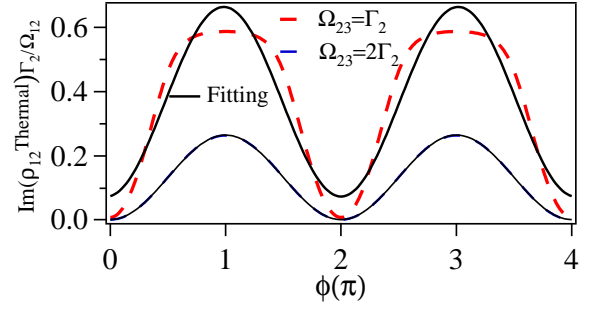}
     \caption{(Color online). Absorption of the probe laser after thermal averaging (Im($\rho^{\textrm{Thermal}}_{12}$)$\Gamma_2/\Omega_{12}$) vs $\phi(\pi)$ at T=1K, with $\delta_{12}=\delta_{23}=\delta_{34}=\delta_{34'}=\delta_{45}=\delta_{4'5'}=\delta_{56}=\delta_{5'6'}=\delta_{36}=\delta_{36'}=0$, $|\Omega^{\textrm{ref}}|=\sqrt{2}\Gamma_2$ and $|\Omega^{\textrm{unk}}|=\sqrt{2} $$\times$0.1$\Gamma_2$  .}
      \label{Abs_vs_Phase}
   \end{center}
\end{figure}

Next, we study the probe absorption after thermal averaging vs the phase $\phi$ with all the detunings at zero. From the plot shown with the red dashed line in Fig. \ref{Abs_vs_Phase}, we observe more than 95$\%$ change in the probe absorption for the change of phase from $0$ to $\pi$ for $|\Omega^{\textrm{unk}}|=\sqrt{2} \times 0.1\Gamma_2$ and for the input parameters i.e. $\Omega_{23}=\Gamma_2$ and $|\Omega^{\textrm{ref}}|=\sqrt{2}\Gamma_2$.   
The numerical data shown by dotted red curve is fitted by a function A+Bsin(f$\phi$+$\theta$), where A, B, f and $\theta$ are kept as free parameters, the fitting is shown with a black curve in Fig. \ref{Abs_vs_Phase}. This shows a strong deviation from the sinusoidal behavior. In order to have sinusoidal behavior we increase the $\Omega_{23}$ to 2$\Gamma_2$ as shown with dotted blue trace.

\begin{figure}
   \begin{center}
      \includegraphics[width =\linewidth]{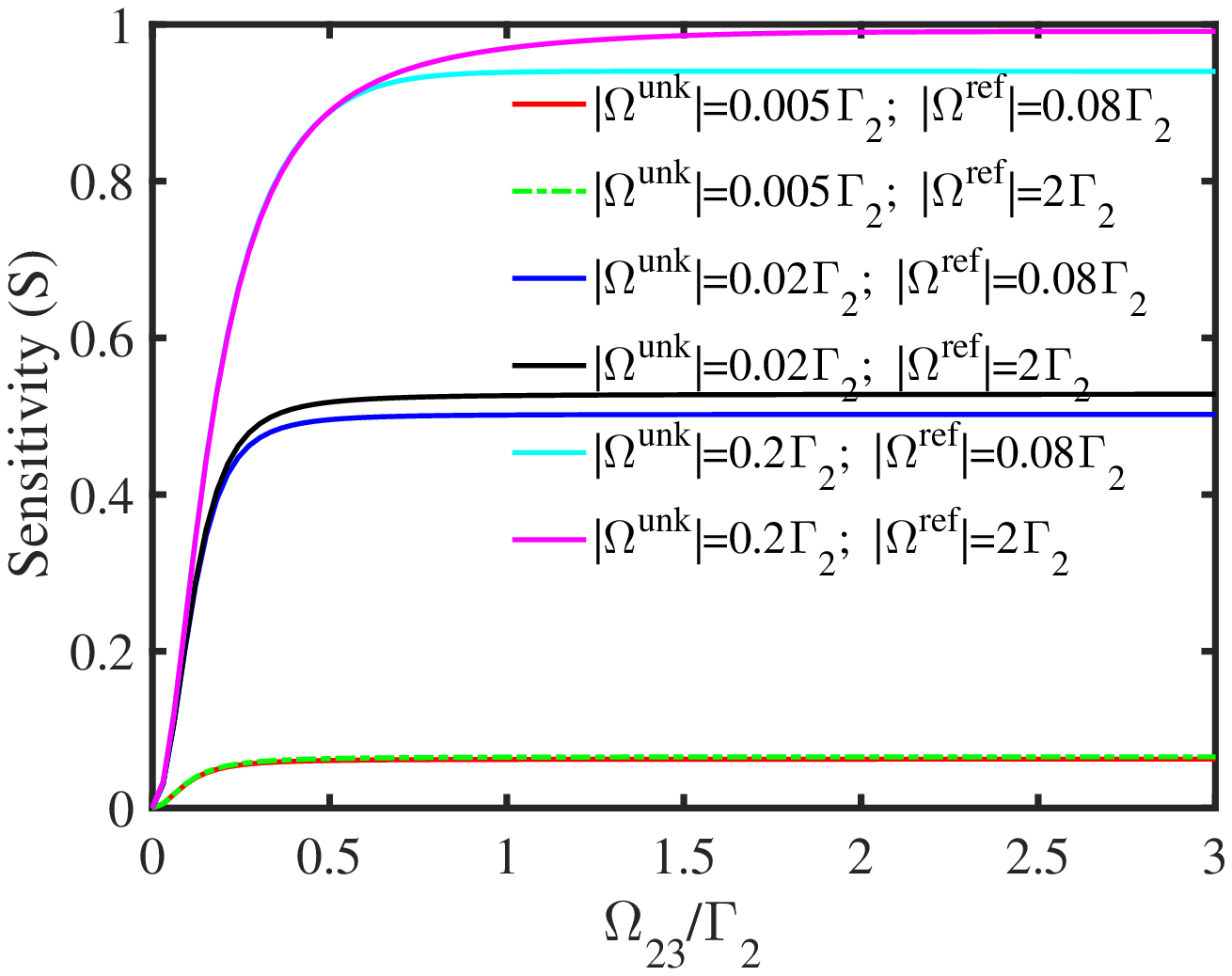}
      \caption{(Color online). $S$ vs $\Omega_{23}/\Gamma_2$ with $\delta_{12}=\delta_{23}=\delta_{34}=\delta_{34'}=\delta_{45}=\delta_{4'5'}=\delta_{56}=\delta_{5'6'}=\delta_{36}=\delta_{36'}=0$.}
      \label{Sensitivity_vs_Omega_23}
   \end{center}
\end{figure}

\begin{figure}
   \begin{center}
      \includegraphics[width =\linewidth]{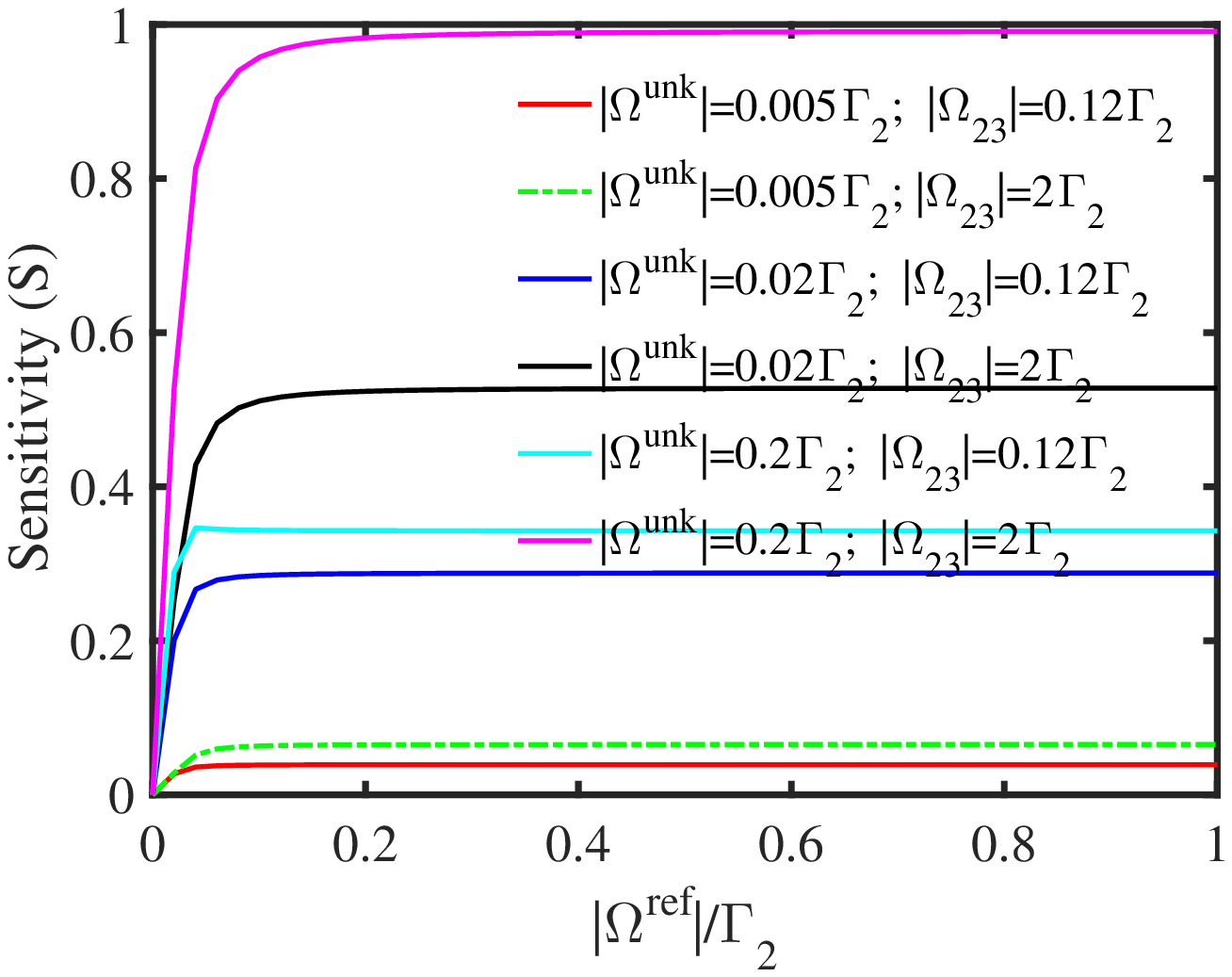}
      \caption{(Color online). $S$ vs $|\Omega^{\textrm{ref}}|/\Gamma_2$ with $\delta_{12}=\delta_{23}=\delta_{34}=\delta_{34'}=\delta_{45}=\delta_{4'5'}=\delta_{56}=\delta_{5'6'}=\delta_{36}=\delta_{36'}=0$.}
      \label{Sensitivity_vs_Omega_ref}
   \end{center}
\end{figure}
To measure the phase and amplitude/strength of the unknown MW field, we define a quantity,  $S$=Im[$\rho^{\textrm{Thermal}}_{12}(\phi=0)-\rho^{\textrm{Thermal}}_{12}(\phi=\pi)$]/Im[$\rho^{\textrm{Thermal}}_{12}(\phi=0)+\rho^{\textrm{Thermal}}_{12}(\phi=\pi)$]. We plot $S$ for different values of $|\Omega^{\textrm{unk}}|$ as a function of the input parameters, $\Omega_{23}$ and  $|\Omega^{\textrm{ref}}|$ in Fig. \ref{Sensitivity_vs_Omega_23} and Fig. \ref{Sensitivity_vs_Omega_ref} respectively. From the figures it is clear that the sensitivity $S$ increases with $\Omega_{23}$ and  $|\Omega^{\textrm{ref}}|$ and then it saturates. The saturation points on $\Omega_{23}$ and  $|\Omega^{\textrm{ref}}|$  increase with increment of $|\Omega^{\textrm{unk}}|$.

\begin{figure}
   \begin{center}
      \includegraphics[width =\linewidth]{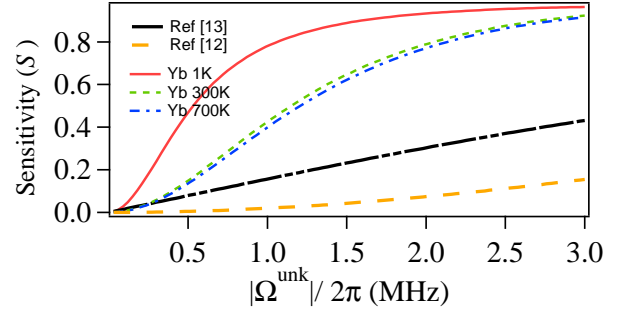}
      \caption{(Color online). (a) $S$  for various system vs $|\Omega^{\textrm{unk}}|/2\pi($MHz$)$ with $\delta_{12}=\delta_{23}=\delta_{34}=\delta_{34'}=\delta_{45}=\delta_{4'5'}=\delta_{56}=\delta_{5'6'}=\delta_{36}=\delta_{36'}=0$ and optimized Rabi frequencies of control fields in individual case.}
      \label{Sensitivity_vs_Omega_unk_comp}
   \end{center}
\end{figure}
\begin{figure}
   \begin{center}
      \includegraphics[width =\linewidth]{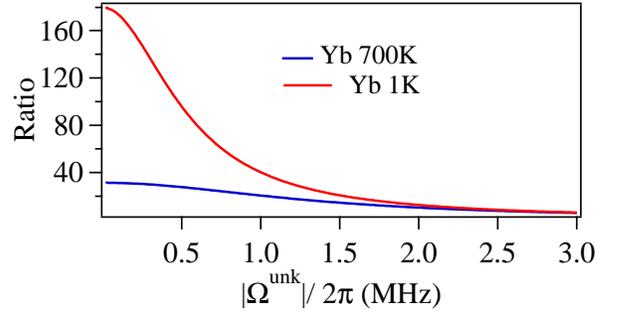}
      \caption{(Color online).Ratio of the sensitivities between the double loopy ladder system and the system in Ref. \cite{SSK12} vs $|\Omega^{\textrm{unk}}|/2\pi($MHz$)$ with $\delta_{12}=\delta_{23}=\delta_{34}=\delta_{34'}=\delta_{45}=\delta_{4'5'}=\delta_{56}=\delta_{5'6'}=\delta_{36}=\delta_{36'}=0$. }
      \label{Ratio_vs_Omega_unk}
   \end{center}
\end{figure}

We also compare the strength sensitivity for the MW field, between the previously studied four-level \cite{SSK12} and six-level loopy \cite{SOP18} ladder systems in Rb with the system studied in this work i.e. the double loopy ladder system in Yb. The sensitivity for the various systems are plotted in Fig. \ref{Sensitivity_vs_Omega_unk_comp}. The sensitivity of the double loopy ladder system in Yb is much higher than the four-level \cite{SSK12} system. This is due to the fact that the effect of small $|\Omega^{\textrm{unk}}|$ gets amplified by the strong control $|\Omega^{\textrm{ref}}|$ as both appears in multiplication inside the \textbf{int} and \textbf{int$'$} terms in Eq. 8 and Eq. \ref{anasol} respectively. The ratio of the sensitivities between the two systems vs $|\Omega^{\textrm{unk}}|$ at two different temperature is plotted in the Fig. \ref {Ratio_vs_Omega_unk}. 

\begin{figure}
   \begin{center}
      \includegraphics[width =\linewidth]{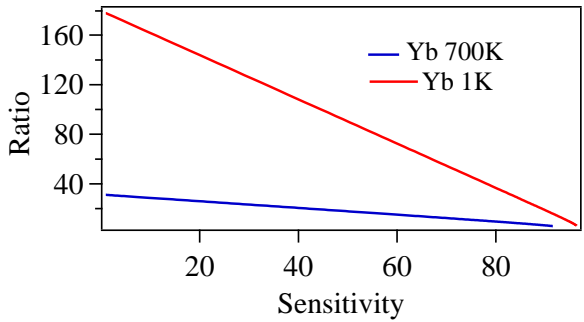}
      \caption{(Color online). Ratio of the sensitivities between the double loopy ladder system and the system in Ref. \cite{SSK12} vs sensitivity of the double loopy ladder system. }
      \label{Ratio_vs_Sensitivity}
   \end{center}
\end{figure}

We also plot the ratio of sensitivities between the double loopy ladder system to the system in Ref. \cite{SSK12} vs sensitivity of double loopy ladder system which gives the information about the possibility of the detection of $|\Omega^{\textrm{unk}}|$. This is an important plot because there is a possibility that the ratio is large but cannot be detected by the double loopy ladder system in Yb. This detection of the  Sensitivity, $S$ upto 1$\%$ is very much possible by using locking detection. At this value, the ratio of the sensitivities between the double loopy ladder system in Yb and the four-level system in Rb at 1~K  and 700~K \cite{SSK12} will be around $200$  and 40 respectively as shown in Fig. \ref{Ratio_vs_Sensitivity}.

The higher sensitivity of this system with respect to the six-level loopy ladder system\cite{SOP18} in Rb  is due to very small mismatch of Doppler shift for the probe at 398~nm and the control laser 395~nm as compared to the probe at 780~nm and the control laser at 480~nm used in Rb\cite{SOP18}. 

%Finally one more important point is that, for the six-level loopy ladder system the MW field $\Omega^{\textrm{unk}}$ can be detected by just varying the phase of the reference MW field, while in the case of the four-level system \cite{SSK13} we need to insert and remove MW mechanical shield.

\section{Conclusions}
In conclusion, we theoretically study a scheme to develop the single reference atomic based MW interferometry in Yb using Rydberg states instead of three reference MW fields as compared to our previous study. This is based upon the interference between the two sets of sub-systems causing EITATA. The interference is either constructive or destructive depending upon the phase of the unknown MW field w.r.t reference MW field. Thereby, this system provides a great opportunity to characterize the MW electric fields completely including the propagation direction and the wavefront. Further, this system is \textbf{two orders} of magnitude more sensitivity to the field strength as compared to previous experimental demonstration on MW electrometry. The bandwidth of the atomic based interferometry ranges from MHz, GHz upto THz. This work will be quite useful in the areas of communications particularly in active radar technologies and synthetic aperture radar interferometry.

\section{Acknowledgment}
K.P. would like to acknowledge the funding from SERB of grant No. ECR/2017/000781 and discussion with David Wilkowski at CQT NTU. D.S. would like to acknowledge the financial support from the Council of Scientific and Industrial Research, India.

\bibliography{eitrefsall}

\end{document}